\documentclass[aps,prd,twocolumn,nofootinbib,superscriptaddress]{revtex4-2}
\usepackage{graphicx}
\usepackage{dcolumn}
\usepackage{bm}
\usepackage[dvipsnames]{xcolor}
\usepackage{amssymb} 
\usepackage{amsmath}
\usepackage{booktabs}

\usepackage{tcolorbox}
\usepackage{siunitx}
\usepackage{caption}
\usepackage{subcaption}
\usepackage{amssymb,amsmath,amsfonts,amsthm}

\usepackage{IEEEtrantools}
\usepackage{appendix}

\usepackage{hyperref}
\hypersetup{pagebackref=false,
  hyperindex=true,
  citecolor=blue,
  colorlinks=true,
  linkcolor=blue,
urlcolor=blue}

\usepackage[capitalize]{cleveref}
\usepackage{pifont}
\usepackage{xcolor}
\newcommand{\cmark}{\textcolor{green!60!black}{\ding{51}}} 
\newcommand{\xmark}{\textcolor{red!60!black}{\ding{55}}} 

\usepackage{orcidlink}

\begin{document}

\title{Assessing signal cross talk between extreme-mass-ratio inspirals and Galactic binaries in LISA data}
\author{Sviatoslav Khukhlaev\,\orcidlink{0009-0000-6749-7232}}
\email{s52skhuk@uni-bonn.de}
\affiliation{Argelander Institut für Astronomie, Universität Bonn, Auf dem Hügel 71, 53121 Bonn, Germany}
\author{Stanislav Babak\,\orcidlink{0000-0001-7469-4250}}
\affiliation{
 Astroparticule et Cosmologie, CNRS, Universit\'e Paris Cit\'e, F-75013 Paris, France
}

\date{\today}

\begin{abstract}

The future space-based gravitational wave observatory, the Laser Interferometer Space Antenna, is expected to observe between 1--1000s extreme mass-ratio inspirals (EMRIs) per year. Due to the simultaneous presence of other gravitational wave signals in the data, it can be challenging to detect EMRIs and accurately estimate their parameters. In this work, we investigate the interaction between a gravitational signal from an EMRI and millions of signals from inspiralling Galactic white dwarf binaries. We demonstrate that bright Galactic binaries can contaminate the detection and characterization of EMRIs. We perform Bayesian inference of EMRI parameters after removing resolvable Galactic binaries and confirm an accuracy comparable to that expected in Gaussian noise.

\end{abstract}

\maketitle

\section{Introduction}
\label{sec:Intro}

Laser Interferometer Space Antenna (LISA) is a space-based gravitational wave observatory planned for launch around 2035. It will consist of three identical spacecraft orbiting the Sun in a triangular formation, with each side of the triangle measuring approximately 2.5 million kilometers. LISA will observe gravitational wave (GW) signals in the frequency range from $0.1$ mHz to $1$ Hz~\cite{LISA:2017, LISA:2024hlh}.

One of the most important sources for LISA is extreme mass-ratio inspirals (EMRIs). An EMRI is a binary system consisting of a massive black hole (MBH) with mass $M > 10^5\,M_\odot$ and a compact stellar-mass object orbiting it. The compact object (CO) undergoes $\sim10^6$ revolutions in the vicinity of the central MBH before the plunge, emitting a detectable GW signal throughout the inspiral~\cite{Amaro-Seoane:2007osp}. LISA is expected to observe between 1--1000s EMRIs per year~\cite{Babak:2017tow}.

Accurate estimation of EMRI parameters is required for a range of astrophysical studies, as it can provide insights into the mechanisms of EMRI formation, test the nature of black holes, and offer an independent measurement of the Hubble constant~\cite{Amaro-Seoane:2007osp}. Moreover, recent research suggests that it may be possible to probe the physics of accretion onto MBHs if a compact object is embedded in a gaseous disk~\cite{Speri:2022upm}.

Although the large number of orbital cycles should, in principle, allow very precise parameter estimation for EMRIs, in practice, this is a complex task due to the simultaneous presence of other strong signals in the data. The problem of simultaneously detecting and characterizing all GW signals and noise in LISA data analysis is called the ``global fit.'' Several successful attempts at the preliminary global fit have been carried out on simulated LISA data, codenamed “Sangria”~\cite{Deng:2025wgk, Littenberg:2023xpl, Katz:2024oqg, Strub:2024kbe}, which was released as part of the second round of the LISA Data Challenges (LDC).

The LDC comprises several challenges, each
designed to thoroughly test and refine data analysis techniques for LISA. The first set of challenges (LDC1) consists of six scenarios, each with only one type of source present. In contrast, LDC2a (“Sangria”)~{\cite{Sangria}} was created to enable early attempts at the global fit. The Sangria dataset contains simulated instrumental noise, a population of inspiralling Galactic white dwarf binaries (GBs), and a population of merging MBHs; see \cite{Deng:2025wgk} for details. Current global fit pipelines can recover GW signals from merging massive black holes and resolve more than 6000 GBs~\cite{Deng:2025wgk, Littenberg:2023xpl, Katz:2024oqg, Strub:2024kbe}. Most GBs ($>20$ million) form a stochastic foreground signal that dominates over the instrumental noise in the frequency range $0.3$–$3$ mHz~\cite{Ruiter:2007xx}.  

EMRIs have not yet been included in global fit studies, mainly because these sources are difficult to detect even in Gaussian instrumental noise~\cite{Gair:2004iv, Babak:2017tow, Chua:2021aah, Strub:2025dfs}. Even sources with high signal-to-noise ratio (SNR) are usually instantaneously quiet sources: to detect them, it is necessary to integrate over thousands to millions of cycles and over multiple harmonics of the EMRI signal. The motion of a compact stellar-mass object around an MBH can be decomposed into azimuthal ($\phi$), radial ($\chi$), and polar ($\theta$) motion with non-commensurate frequencies. In a Keplerian orbit, all three frequencies are equal, whereas in General Relativity there is precession of the apoapsis and precession of the orbital plane around the MBH spin (due to spin–orbit coupling). These frequencies slowly evolve under radiation reaction. The total EMRI signal is therefore a “chorus” of multiple harmonics of these fundamental frequencies. Here we give an oversimplified description, focusing on the GW features relevant for our analysis; more detailed discussions can be found in \cite{Katz:2021yft}.

In this paper, we do not address the problem of EMRI detection; we assume that old methods~\cite{Babak:2009ua, Cornish:2008zd} or more recent approaches~\cite{Strub:2025dfs} work well. Our main objective is to investigate and quantify the interaction (``cross talk'') between EMRIs and other GW sources, in particular GBs. Merging MBHs (massive black hole binaries, MBHBs) produce the strongest GW signals in LISA data, but most of their SNR accumulates in the few weeks prior to merger, making them effectively transient sources. Current rate estimates for MBH mergers are uncertain but not very high, ranging between 1--100s per year~\cite{LISA:2022yao}. We assume that these signals can be accurately characterized, so that residuals after subtraction contain a negligible amount of GW power associated with them. We also assume that their recovery is not affected by the presence of EMRIs; therefore, we focus only on the superposition of EMRIs with millions of GBs. 

GW signals from a vast majority of GBs are almost monochromatic and persist for the entire LISA mission. Since EMRI harmonics evolve slowly over time, certain combinations of strong GB signals can resemble an EMRI and bias parameter estimation, and we explore this possibility here. We note that a handful of potentially detectable binaries can be eccentric~\cite{Willems:2007xe, Kremer:2018tzm} and may further accentuate the EMRI-GB confusion problem. However, such sources are not included in Sangria and are therefore neglected here.

Previous theoretical work~\cite{Racine:2007gv} suggested that the superposition of unresolved GBs would likely effectively form a Gaussian background for EMRI searches, with only small non-Gaussian corrections, but this has not been tested with simulated data. The main result of this paper is that strong GB sources can indeed mimic EMRI signals, leading to severe parameter estimation biases and false detections. We confirm that the unresolved (and weak) GB component of the Galaxy does not affect EMRI parameter estimation.  

Throughout the paper, we use geometrical units $c=G=1$ and Planck 2018 cosmological parameters~\cite{Planck:2018vyg} for redshift calculations.

This paper is structured as follows. In Section~\ref{sec:Model}, we discuss the signal model, the dataset, and the methods used in this work. Then, in Section~\ref{sec:Results}, we present the results of Bayesian inference of the EMRI parameters and address the problem of false detection of EMRI. Finally, in Section~\ref{Sec:Discussion}, we summarize our main results. 

\section{Model}
\label{sec:Model}

As mentioned in the Introduction, we want to simulate LISA data that contain an EMRI signal and a population of GBs. In the next subsections, we give details on the data simulation and design of the experiments.

\subsection{Waveform model for EMRI}

We simulate the EMRI signal using the package \texttt{FastEMRIWaveforms} (\texttt{FEW}) (v1.5.1, July 2023)~\cite{Chua:2020stf, Katz:2021yft, Speri:2023jte, few1}. In particular, we use the \texttt{Pn5AAKWaveform} model. This model is not self-force–accurate, however, it qualitatively reproduces the key features of realistic EMRIs, such as three evolving (fundamental) orbital frequencies, periapsis and Lense–Thirring precession, and multi-harmonic structure~\cite{Chua:2017ujo, Barack:2003fp, Katz:2021yft}.

We note that the newer release of the package (v2.0.0, June 2025)~\cite{Chapman-Bird:2025xtd, few2} introduces faster and more accurate waveforms, which will be useful for follow-up studies.

The LISA response in A and E TDI channels (time-delay interferometry, see~\cite{Tinto:2020fcc}) was applied using interpolations and a GPU implementation in the package \texttt{fastlisaresponse}~\cite{Katz:2022yqe}. We ignore the $T$ channel in this study, as it is dominated by noise in the relevant frequency range.

The TDI time series were transformed to the frequency domain and cropped to the frequency range where the signal is comparable to the noise ($0.6$–$20$ mHz), in order to reduce computation time. The typical waveform evaluation time for one set of parameters was about 6 seconds on an NVIDIA V100 GPU. 

\subsection{Selected EMRI Signal}

The EMRI parameter space is very vast, covering it all is a big task; instead, we want to concentrate on one reference (typical) EMRI waveform. We will investigate it in detail, and at the end try to generalize our findings and comment on the plausible limitations due to this particular choice. 

The detector-frame MBH mass was chosen to be $M\sim6\times10^5\,M_\odot$ based on two criteria:
(i) it lies near the optimal mass for EMRI detection (largest reach), estimated to be in the $10^5-10^6\,M_\odot$ range~\cite{Babak:2017tow}, and (ii) it maximizes the overlap of the EMRI signal with the GB population in the frequency domain. 

The stellar-mass compact object plunges after $T_{\text{plunge}} = 0.732\,$yr. The total SNR of the signal is about 50 ($\text{SNR}_\text{EMRI} = 49.56$). This is not a particularly strong signal, although it is well above the expected detection threshold of 20~\cite{Babak:2017tow}. Population studies indicate that such a source is typical and falls within the astrophysical range expected for EMRIs (see Fig. 6 of~\cite{Babak:2017tow}). Other parameters are summarized in Table~\ref{table:signal_param}. Note that the masses presented in the table are detector-frame masses, which are related to the astrophysically important source-frame masses as $m_\text{det} = m_\text{source}(1+z)$, where $z$ is the redshift of the source. 

\begin{table}
\centering                          
\begin{tabular}{l|c}        
\specialrule{1pt}{0pt}{0pt} 
\textbf{Parameter} & \textbf{Value} \\ \specialrule{1pt}{0pt}{0pt} 
\textbf{MBH detector-frame mass ($M$)} & $6.087 \times 10^5\,M_\odot$ \\ 
\textbf{CO detector-frame mass ($\mu$)} & $18.06\,M_\odot$  \\ 
\textbf{MBH spin ($a$)} & 0.6 \\  
\textbf{Init. eccentricity ($e_0$)} & 0.3857 \\ 
\textbf{Init. semi-latus rectum ($p_0$)} & $10.75\,M$ ($0.065\,$au)\\ 
\textbf{Init. cos of inclination ($x_0$)} & 0.7  \\ 
\textbf{Luminosity distance ($D_L$)} & $3.59\,$Gpc \\  
\textbf{Polar spin angle ($q_K$)}  & 0.9433 \\   
\textbf{Azimuthal spin angle ($\phi_K$)} & 1.726 \\  
\textbf{Polar sky angle ($q_S$)} & 2.176 \\  
\textbf{Azimuthal sky angle ($\phi_S$)} & 0.7101 \\ \hline 
Plunge time ($T_\text{plunge}$) & $0.732\,$yr \\
Redshift ($z$) & 0.594 \\
Time step ($dt$) & $5\,$s  \\ 
Init. azimuthal phase ($\phi_0$) & 0 \\  
Init. polar phase ($\theta_0$) & 1.726 \\  
Init. radial phase ($\chi_0$) & 0 \\
\specialrule{1pt}{0pt}{0pt} 
\end{tabular}
\vspace{5pt}
\caption{Reference signal (``true'' point). The 11 parameters inferred in the parameter estimation are listed in bold.}
\label{table:signal_param}
\end{table}

\subsection{Simulated LISA data}
We adopted the publicly available ``Sangria'' dataset in this work~\cite{Sangria}. The dataset is one year long and contains a population of approximately 30 million Galactic binaries and 15 merging MBHBs. A detailed description of the data is provided in~\cite{Deng:2025wgk}.

\begin{figure}
\centering
\includegraphics[width=\hsize]{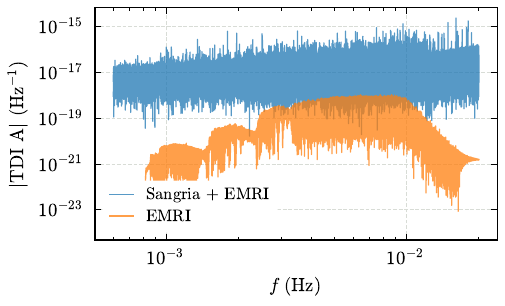}
\caption{EMRI and total Sangria signal (excluding MBHBs) vs frequency (log scale).}
\label{fig:emri_sangria}
\end{figure}

As discussed in the Introduction, we have removed the signals from MBHBs and added the EMRI signal described in the previous subsection. The resulting data in the frequency domain is shown in Fig.~\ref{fig:emri_sangria}. The full dataset for the A-TDI channel is shown in blue, with the EMRI signal overlaid in orange.

Of the 30 million GBs present in Sangria, a few thousand are individually resolvable. In this study, our objective was to test the influence of both the resolvable and unresolved subpopulations of GBs on the recovery of the EMRI signal. We use the results presented in~\cite{Deng:2025wgk} to identify the sources considered as ``resolvable.''

\subsection{Methods}

We adopted a Bayesian approach for most of the work in this paper. We infer the posterior distribution
of EMRI parameters using parallel tempering Markov Chain Monte Carlo (MCMC), with a specific implementation \texttt{m3c2} described in~\cite{Falxa:2022yrm} and publicly available at \url{https://gitlab.in2p3.fr/lisa-apc/m3c2}. These techniques are commonly utilized in LISA data analysis~\cite{Cornish:2005qw}.

We assume that the data $d$, after correct subtraction of the EMRI signal, consists of Gaussian noise, and we use the corresponding likelihood function

\begin{equation}
p(d|\boldsymbol{\theta}, \mathcal{H}) \propto e^{-\frac{1}{2}(d - h(\boldsymbol{\theta})|d - h(\boldsymbol{\theta}))}\,,
\end{equation}

where $\mathcal{H}$ is the assumed model and $h(\boldsymbol{\theta})$ is the simulated waveform, which depends on the 11 parameters listed in bold in Table~\ref{table:signal_param}. The inner product is defined as the matched filtering in the frequency domain:

\begin{equation}
(a| b) = 4 \Re \int\limits_{0}^{+\infty} \dfrac{a(f) b^*(f)}{S_n (f)} df\,,
\end{equation}

where $S_n(f)$ is the one-sided noise power spectral density (PSD).

We introduce two additional quantities that will be useful in our discussion. The matched-filtering SNR, $\rho$,
is defined as

\begin{equation}
\rho = \frac{(h|d)}{\sqrt{(h|h)}}
\end{equation}

and depends on the noise realization. In the high-SNR limit, it approaches the optimal-SNR, $\sqrt{(h|h)}$. Most resolvable EMRIs are expected to lie near the adopted threshold $\rho_\text{thresh} \approx 20$~\cite{Babak:2017tow}, where the influence of noise is not negligible; therefore, we will often quote the matched-filtering SNR.

Another useful quantity is the overlap:

\begin{equation}
\text{Overlap}(a, b) = \frac{(a|b)}{\sqrt{(a|a)(b|b)}}\,,
\end{equation}

which indicates the similarity between two signals, independently of their strength. 

\subsection{Setups of experiments}
\label{sec:setups_experiments}

We have performed several investigations, and for each we used slightly different data and models. We first describe their purpose verbally and then summarize in the table.  

\begin{table*}[tbh]
\centering                          
\begin{tabular}{c|c|c|c|c}        
\specialrule{1pt}{0pt}{0pt} 
\textbf{Run Name} & \textbf{EMRI} & \textbf{Resolvable GBs} & \textbf{Unresolvable GBs} & \textbf{Instrumental noise}\\ 
\specialrule{1pt}{0pt}{0pt} 
Noiseless run & \cmark & \xmark & \xmark & \xmark \\ 
Full-Galaxy run & \cmark & \cmark & \cmark & \cmark \\ 
Foreground GBs run & \cmark & \xmark & \cmark & \cmark \\
Resolvable GBs run & \xmark & \cmark & \xmark & \xmark \\ 
Stochastic run & \xmark & \xmark & \cmark & \cmark \\ 
\specialrule{1pt}{0pt}{0pt} 
\end{tabular}
\vspace{5pt}
\caption{Summary of investigations. Symbols: \cmark\ = included; \xmark\ = not included.}
\label{table:runs}
\end{table*}

\begin{enumerate}
    \item ``Noiseless'' run. We perform Bayesian inference of the EMRI parameters in the noiseless data, which consists only of the ``true'' EMRI, while preserving the SNR of the signal (through $S_n(f)$ in the matched filtering). This exercise should give us a clean theoretical posterior which will be used as a reference for further comparisons. The true parameters point should lie within a high-level credible interval and correspond to the maximum likelihood. 

    \item ``Full-Galaxy'' run. As described above, we add EMRI signal into Sangria dataset after removing the GW signals from the merging MBHBs. We keep all $\approx 30$ mln GBs, so the data contains instrumental noise, a population of GBs and one EMRI signal. We will infer the parameters of EMRI using parallel tempering MCMC seeded at the true parameters. The main purpose is to obtain the posterior on the EMRI parameters in the presence of a Galactic population. 

    \item ``Foreground GBs'' run. We investigate separately the effect of resolvable and unresolved components of Galactic population on EMRI inference. In this part, we remove all resolvable GBs and repeat the previous investigation on the data that contains instrumental noise, unresolved (stochastic) GB foreground, and one EMRI signal. This approach reflects the current strategy of the global-fit framework, where the EMRI detection module operates on data cleaned of detected MBHBs and GBs (see~\cite{Deng:2025wgk} for the description of modules).

    \item ``Resolvable GBs'' run. Next, we construct the data \emph{only} from resolvable GBs and perform an EMRI inference. Here, we investigate the possibility of a false EMRI detection conspired by superposition of strong signals from GBs. We run three independent chains seeded at three different locations in the parameter space. One chain is seeded at the parameters of the reference EMRI signal. 

    \item ``Stochastic'' run. Finally, we want to investigate the possibility of false detection of the EMRI signal in data that contain instrumental noise and the \emph{unresolved} stochastic GB foreground. Like in the previous investigation (``Resolvable GBs''), no EMRI signal is present in the data.  
\end{enumerate}

All investigations are summarized in Table~\ref{table:runs}.

We always used uniform priors spanning 40\% symmetrically around both masses and 50\% around the luminosity distance. The MBH spin $a$ was uniform on $[0.1,0.9]$; the initial semi-latus rectum $p_0$ on $[10M,12M]$; the initial eccentricity $e_0$ on $[0.05,0.5]$; and the initial cosine of inclination $x_0$ on $[-0.9,0.9]$. Spin and sky angles had independent isotropic priors on the unit sphere: $(\cos q_{K},\phi_{K}),(\cos q_{S},\phi_{S})\sim\mathcal{U}([-1,1]\times[0,2\pi))$.

In all investigations, we assumed a known PSD, which is a combination of instrumental and unresolved GB foreground noise. In that sense, it implicitly assumes successful subtraction of resolvable GBs. The analytical noise model (named ``sangria'') is given in~\cite{LDC} and derived using the method described in~\cite{Karnesis:2021tsh}. We stress that this is not a PSD estimated in the global fit, but rather an analytical approximation, which is nevertheless close to the PSD estimated in~\cite{Deng:2025wgk}.

Note that the Galactic foreground varies in time due to the motion of LISA, forming cyclo-stationary noise~\cite{Digman:2022jmp}. We have ignored these variations and used the PSD averaged over a one-year period. This implies that our results are somewhat conservative in the size of credible intervals, but the main findings remain valid.

The convergence of the MCMC was monitored visually using corner and trace plots and through an effective number of samples. 

\section{Results}
\label{sec:Results}

In this section, we present results for all investigations mentioned above. The ``Noiseless'' results are used as reference to understand the contribution of the noise and resolvable sources.

\subsection{Influence of the full galaxy}

We start by examining the influence of all GBs present in the Sangria dataset which corresponds to the ``Full-Galaxy'' run in Table~\ref{table:runs}. 

\begin{figure}[h!]
   \centering
   \includegraphics[width=\hsize]{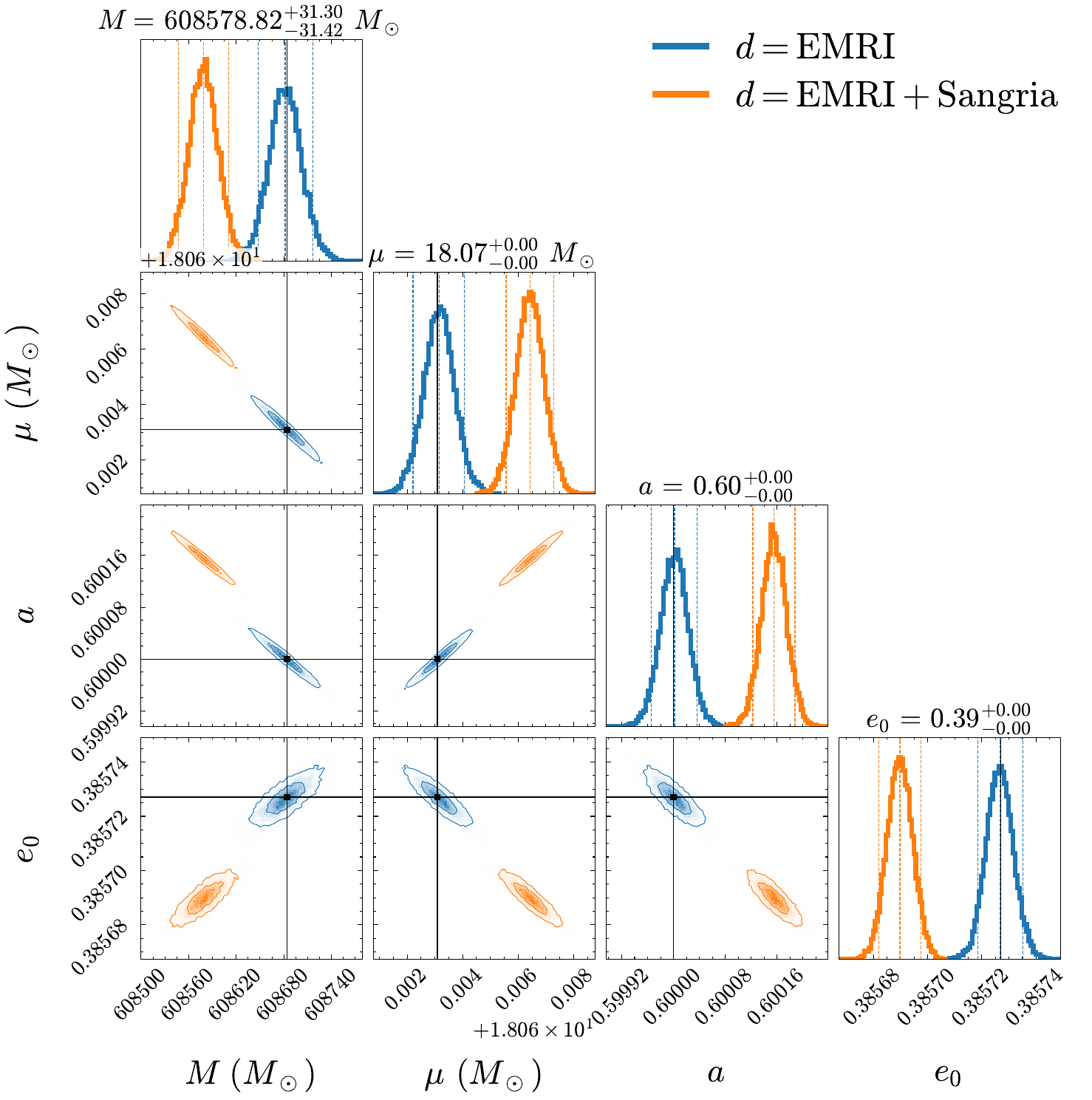}
      \caption{Influence of the full population of Galactic binaries on EMRI parameter estimation. Corner plot shows Bayesian inference of the four most important binary parameters. The true point is marked in black. Contour lines mark the 50\% and 90\% credible regions. The posteriors for the noiseless run are given in blue for comparison.
              }
         \label{fig:main_mcmc}
\end{figure}

The posteriors for the most astrophysically relevant parameters (MBH mass, compact object mass, MBH spin, and initial eccentricity)  are given in Fig.~\ref{fig:main_mcmc} as orange contours corresponding to 50\% and 90\% credible regions. The full plot for 11 parameters can be seen in Appendix~\ref{appendix}. The noiseless results are shown as blue contours for comparison. We observe a statistically significant bias in parameter estimation, although the covariance ellipses appear similar. 

Despite the bias being small in absolute terms (of order 0.01\% for key binary parameters), the fact that the true point lies outside the 90\% credible region can, for example, lead to biased tests of General Relativity. For future reference, we denote the set of maximum likelihood parameters found in the ``Full-Galaxy'' run as $\boldsymbol{\theta}_\text{Sangria}$. 

It is important to note that we have seeded the MCMC at the true parameter set and only performed local explorations around it. This explains the unimodal structure of the posterior, whereas we normally expect strong (remote) secondary maxima~\cite{Chua:2021aah}. 
 
The following subsections are dedicated to understanding these results.

\subsection{Influence of resolvable GB signals}

We divided the total Galactic population into sources identified as resolvable in~\cite{Deng:2025wgk} and the remaining (unresolvable) ones. In total, we have about 9500 resolvable sources. Most of them have SNRs between 8 and 20; however, in some cases, we encounter very strong WD binaries with SNRs above a few hundred. In this subsection, we show that these resolvable sources strongly influence the search and parameter estimation of EMRI signals. For this purpose, we analyze data consisting only(!) of GW signals from resolvable GBs. Note that averaging the cyclo-stationary GB foreground in the analytical PSD model might also introduce a bias. In our example, however, no such effect is observed (see the “Noiseless” results).

\begin{figure}[h!]
   \centering
   \includegraphics[width=\hsize]{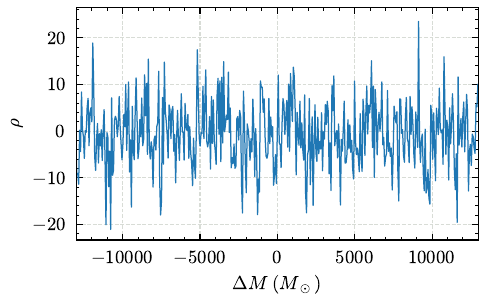}
      \caption{$\rho=(h^\text{EMRI}(\boldsymbol{\theta})| h_\text{resolvable}) / \text{SNR}_\text{EMRI}$ vs $\Delta M$, where $\boldsymbol{\theta} = (M_\text{true} + \Delta M, \mu_\text{true}, \dots)$. The matched filtering SNR does not show a clear dependence on the MBH mass and resembles a popcorn-like noise. Similar behavior is observed for other parameters.
              }
         \label{fig:variations_M}
\end{figure}

First, we evaluated the matched filtering SNR between our EMRI signal and the data, $\rho = (h^\text{EMRI}|h_\text{resolvable}) / \text{SNR}_\text{EMRI}$, near the true EMRI parameters. In Fig.~\ref{fig:variations_M}, we show how $\rho$ varies as we move along the MBH mass (shifting left and right from the true value by $\Delta M$). We observe that (i) the matched-filtering SNR can easily reach $\rho \sim 20$, and (ii) there are multiple strong outliers. Similar behavior of the matched-filtering SNR is observed when varying other parameters. 

\begin{figure}[h!]
   \centering
   \includegraphics[width=\hsize]{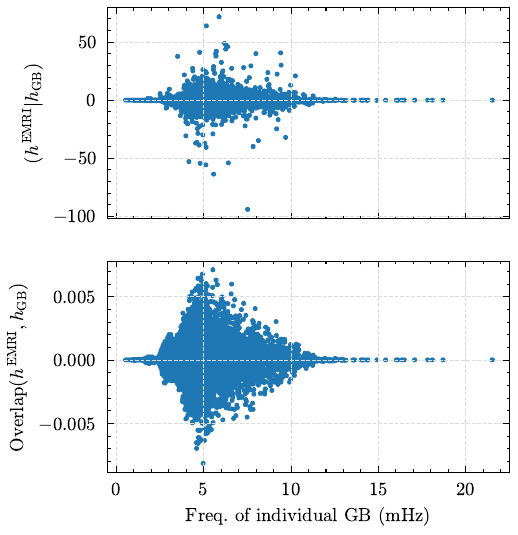}
      \caption{Top: $(h^\text{EMRI}|h_\text{GB})$; Bottom: Overlap$(h^\text{EMRI},h_\text{GB})$ for individual resolvable GBs plotted against each GB's frequency. The EMRI waveform is computed at $\boldsymbol{\theta}_\text{Sangria}$.
              }
         \label{fig:individual_overlap}
\end{figure}

To further understand where this significant matched filtering SNR comes from, we computed the inner product between the EMRI signal and each individual resolvable GB, shown in the top panel of Fig.~\ref{fig:individual_overlap}. We observe that the inner product with individual GBs can reach values as high as 70, translating to the matched filtering SNR $\rho\sim1.4$, while the overlap, plotted in the bottom panel of Fig.~\ref{fig:individual_overlap}, shows very low values. This implies that individual GBs are not similar to the EMRI signal, and it is their combined contribution that can produce a significant matched-filtering SNR that was observed in Fig.~\ref{fig:variations_M}.

\begin{figure}[h!]
   \centering
   \includegraphics[width=\hsize]{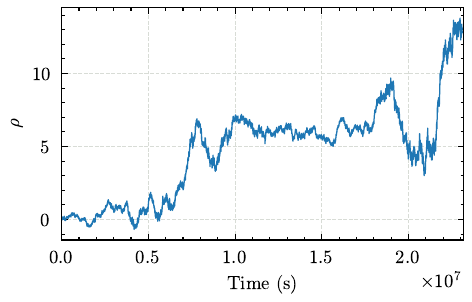}
   \includegraphics[width=\hsize]{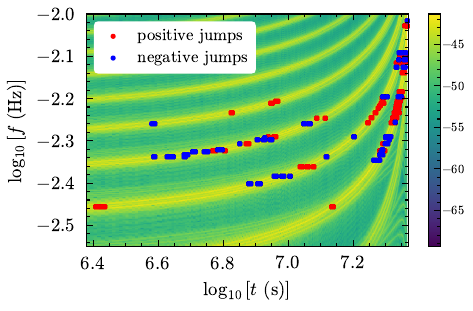}
      \caption{Top: Cumulative matched filtering SNR, $\rho(t)$, of EMRI signal with parameters $\boldsymbol{\theta}_\text{Sangria}$ and all resolvable GBs.
      Bottom: Time-frequency plot for the $\boldsymbol{\theta}_\text{Sangria}$ point. The spectrogram demonstrates only TDI A and is displayed on a log scale. The blue and red points mark detected sharp jumps in the cumulative inner product of EMRI with individual resolvable GBs.
              }
         \label{fig:time_freq_max_like}
\end{figure}

Using the ``Resolvable GBs'' data, we computed the cumulative matched-filtering SNR, $\rho(t)$, for the EMRI signal with parameters $\boldsymbol{\theta}_\text{Sangria}$, corresponding to the orange contours in Fig.~\ref{fig:main_mcmc}. The result is shown in the top panel of Fig.~\ref{fig:time_freq_max_like}. Visible upward and downward jumps correspond to positive (in-phase) and negative (antiphase) interference of EMRI harmonics with GBs, as explained next.

GB signals are monochromatic; they interact with EMRIs when the EMRI harmonics intersect the GB signals in frequency. We further confirm that occasionally a GB not only intersects EMRI harmonics but also stays in phase with them for short periods of time. Such intersections (in-phase or antiphase) produce sharp jumps in the inner product (and correspondingly in the likelihood), and we overlay these jumps on the time-frequency map of the EMRI signal, shown in the bottom panel of Fig.~\ref{fig:time_freq_max_like}. The blue points correspond to drops in $\rho$ at the crossing times, and red points mark frequencies of GBs that add positive jumps in $\rho(t)$. 

This plot confirms that the influence of resolvable GBs is a \emph{net} effect. Both positive and negative jumps are distributed chaotically along the harmonics and do not show any clear pattern, confirming the random nature of the accumulated matched-filtering SNR.

This investigation explains the results of the parameter estimation presented in the previous subsection and shown in Fig.~\ref{fig:main_mcmc}. The chain, while mostly locked onto the simulated EMRI signal in the data, was affected by the combined interaction with bright GBs while exploring the neighborhood in parameter space. This caused a significant bias in the parameters: the chain moved from the true parameters to $\boldsymbol{\theta}_\text{Sangria}$, which corresponds to a significantly higher likelihood solution due to the positive net contribution of bright GBs. The fact that the posterior shape and size were preserved reflects that the template (chain) is still locked onto a few strong harmonics of the EMRI signal, which largely define the posterior.

The strongly non-Gaussian signal from all resolvable GBs significantly impacts not only parameter estimation, but also the search for EMRIs. In fact, performing a primitive search (parallel tempering MCMC, ``Resolvable GBs'' run) shows that we can easily obtain a false detection with SNR $\rho \sim 40$ in different parts of the parameter space. This is one of the main results of our paper. In contrast, previous work demonstrated the detection of EMRIs with $\text{SNR} \sim 20$ in Gaussian noise~\cite{MockLISADataChallengeTaskForce:2009wir}, and this threshold has been used for EMRI detection since then~\cite{Babak:2017tow}. We denote the maximum likelihood point found in the ``Resolvable GBs'' run as $\boldsymbol{\theta_\text{resolvable}}$.

\begin{figure}[h!]
   \centering
   \includegraphics[width=\hsize]{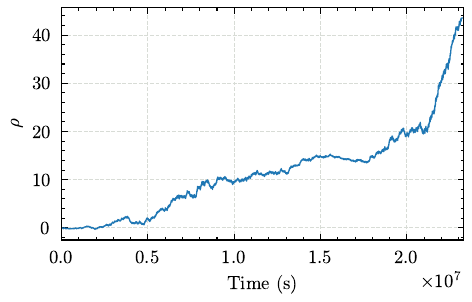}
   \includegraphics[width=\hsize]{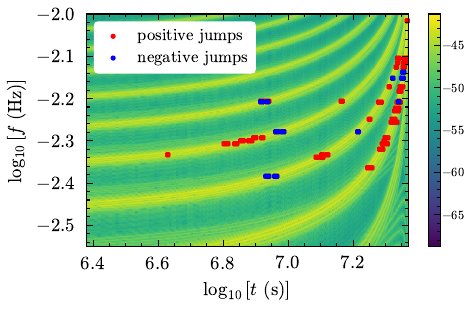}
      \caption{Similar to Fig.\protect{~\ref{fig:time_freq_max_like}} but for the set of parameters 
      $\boldsymbol{\theta}_\text{resolvable}$ corresponding to the false detection with the highest likelihood.
              }
         \label{fig:time_freq_max_like_only_bright}
\end{figure}

The cumulative $\rho(t)$ for the $\boldsymbol{\theta_\text{resolvable}}$ point is shown in the top panel of Fig.~\ref{fig:time_freq_max_like_only_bright}. Similarly to Fig.~\ref{fig:time_freq_max_like}, we overlay the resolvable GBs that contributed the most significantly to the net SNR in the time-frequency plot shown in the bottom panel of Fig.~\ref{fig:time_freq_max_like_only_bright}. The search identified a fake EMRI signal with a relatively smooth accumulation of $\rho(t)$ and only a few negative jumps (shown in blue).

As expected, the most significant jumps cluster around the strongest EMRI harmonics and often at high frequencies, where both the EMRI and the GB signals have larger amplitudes. This further confirms the necessity of removing \emph{all} resolvable GBs, not only a few bright ones, before performing EMRI searches and parameter estimation.

\subsection{Parameter estimation in the presence of unresolved Galactic population}
\label{sec:unresolved_mcmc}

In this part, we study the interaction of the unresolved Galactic population with the EMRI signal; this corresponds to the ``Foreground GBs'' run. We added the EMRI signal to the ``Sangria'' data and removed all resolvable GBs. We then repeated the Bayesian inference of EMRI parameters, seeding the chain at the true parameters. 

\begin{figure}[h!]
   \includegraphics[width=\hsize]{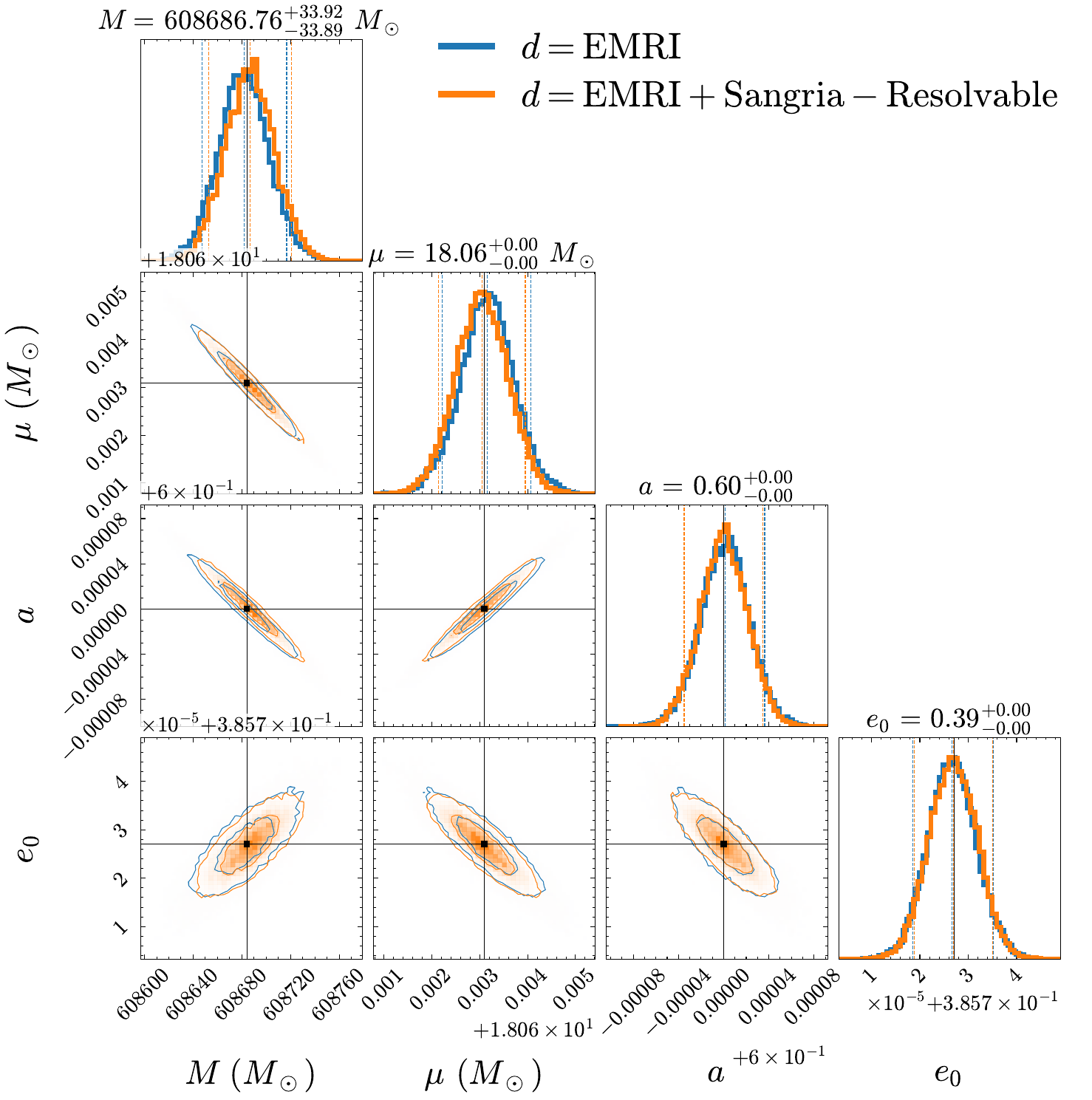}
      \caption{Parameter estimation in the absence of resolvable GB signals. Corner plot shows Bayesian inference of the four most important binary parameters. The true point is marked in black. Contour lines mark the 50\% and 90\% credible regions. The posteriors for the noiseless run are given in blue for comparison.
              }
         \label{fig:without_bright_mcmc}
\end{figure}

The results are presented in Fig.~\ref{fig:without_bright_mcmc}. The orange contours (50\% and 90\% credible regions) represent the inferred posteriors, which we compare to the ``noiseless'' inference shown in blue. We observe excellent agreement, suggesting that the EMRI signal perceives the unresolved component as Gaussian noise.

\begin{figure}[h!]
   \includegraphics[width=\hsize]{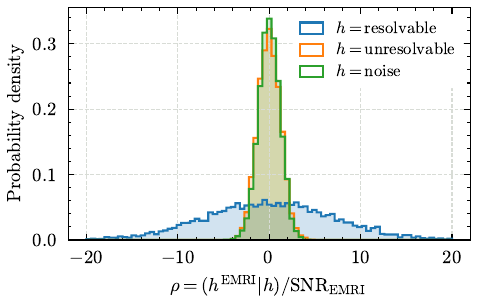}
      \caption{Distributions of the matched filtering SNR $\rho$ for resolvable GBs (see Fig.\protect{~\ref{fig:variations_M}}), unresolved GB foreground and instrumental noise and in the vicinity of the ``true'' parameters.
      }
         \label{fig:histogram_M}
\end{figure}

To further confirm this, we evaluated the matched-filtering SNR $\rho$ using (i) instrumental noise, and (ii) unresolved GB foreground, in the vicinity of the source parameters. The resulting distributions of $\rho$ are shown in Fig.~\ref{fig:histogram_M} as green and orange histograms. For comparison, we have overplotted the distribution of $\rho$ from Fig.~\ref{fig:variations_M}, obtained by matched filtering of the EMRI signal with resolvable GBs. The orange and green histograms are statistically consistent, confirming the Gaussianity of the GB foreground (at least in the vicinity of the EMRI parameters).

\subsection{EMRI response to GBs subtraction}

In the previous subsections, we showed that the cumulative effect of resolvable GBs can bias EMRI parameter estimation. The remaining question is whether this bias is primarily driven by a few very strong GBs or by the combined effect of multiple weaker sources. In this subsection, we study EMRI false detections and parameter estimation as a function of the SNR of resolvable GBs.

We performed three independent Bayesian inferences of EMRI parameters using the same settings as the ``Foreground GBs'' run, but retaining only resolvable GBs above a fixed (artificial) SNR threshold of 100, 50, or 20. We found that visible bias is still present in all cases. However, for the SNR threshold of 20, the bias is smaller, and the maximum-likelihood point lies within the 90\% credible region of the noiseless run.

\begin{figure}[h!]
   \centering
   \includegraphics[width=\hsize]{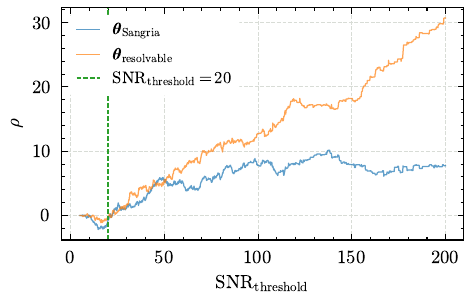}
      \caption{Cumulative matched filtering SNR of EMRI with resolvable GBs as a function GBs strength expressed in terms of their SNR:
      $\rho=(h^\text{EMRI}(\boldsymbol{\theta)}| h_{\text{resolvable, SNR}<\text{SNR}_\text{threshold}}) / \text{SNR}_\text{EMRI}$. 
              }
         \label{fig:snr_threshold}
\end{figure}

Additionally, we investigated false detections by computing the cumulative SNR of an EMRI template against data containing resolvable GBs, similar to Fig.~\ref{fig:time_freq_max_like} and Fig.~\ref{fig:time_freq_max_like_only_bright}. However, in this test, the cumulative matched-filtering SNR, $\rho$, was computed for EMRIs with parameters $\boldsymbol{\theta}_\text{Sangria}$ and $\boldsymbol{\theta}_\text{resolvable}$ as a function of the threshold SNR for the resolvable GBs. The results, presented in Fig.~\ref{fig:snr_threshold}, show that $\rho$ rises sharply when including GBs with $\mathrm{SNR} \gtrsim 20$, indicating that the brightest Galactic binaries are primarily responsible for EMRI false detections.

The fact that most of the bias and false detections in EMRI recovery originate from interactions with bright GBs is reassuring, as it relaxes the requirement on the subtraction of weaker resolvable GBs. However, we argue that EMRIs are ideal laboratories for testing general relativity and studying their astrophysical environments~\cite{Babak:2017tow, Cardenas-Avendano2024}. This means that systematic effects in the global fit must be well controlled, which still requires the removal of all theoretically resolvable GBs. Our results also indicate that the current performance of the global fit~{\cite{Deng:2025wgk}} is already sufficient for reliable EMRI inference (see Fig.~\ref{fig:without_bright_mcmc}).

\subsection{Galactic binaries confusion noise}

Section~\ref{sec:unresolved_mcmc} suggests that the GB foreground is ``seen'' by our EMRI signal as a Gaussian noise component. In this subsection, we further confirm this statement by performing a stochastic search for the EMRI signal in data containing instrumental noise and the GB foreground (the unresolvable component of the Galactic population). This corresponds to the ``Stochastic'' run.

We ran several MCMC chains with parallel tempering, seeded in different parts of the prior, as summarized in Section~\ref{sec:setups_experiments}. We do not claim that this is an exhaustive search or full exploration of the false-alarm probability for EMRI signals. This ``search'' is similar to what was done in the ``Resolvable GBs'' run, where we found false signals with $\rho \sim 40$. The highest matched-filtering SNR (false detection) observed in the ``Stochastic run'' is $\rho = 7.1$. The highest matched-filtering SNR from 20,000 random draws of points from the prior was $\rho = 1.8$.

This is not a formal proof, but it strongly suggests that it is safe to search for EMRIs and to perform Bayesian parameter inference after removing resolvable GB signals.

\section{Summary}
\label{Sec:Discussion}

In this work, we investigated the interaction (``cross talk'') between an EMRI signal and the population of inspiralling Galactic white dwarf binaries in LISA data. We used the simulated ``Sangria'' dataset~\cite{Sangria} and an EMRI signal generated using the \texttt{FEW}~\cite{Chua:2020stf, Katz:2021yft, Speri:2023jte, few1} model.

Our first main conclusion is that the resolvable GB sources (defined as those identified in the global fit in~\cite{Deng:2025wgk}) strongly contaminate
\begin{itemize}
\item the detection of EMRIs, producing false alarms with matched-filtering SNRs as high as $\rho \gtrsim 40$, and
\item Bayesian parameter inference, introducing strong biases that push parameters well outside the expected 90\% credible intervals.
\end{itemize}

Our second main conclusion is that the stochastic GB foreground (after removing resolvable GBs) is ``seen'' by the EMRI signal as Gaussian noise and does not affect EMRI detection and parameter estimation. Both results are important in developing the strategies required for LISA data analysis, especially in science investigations that involve EMRI signals~\cite{LISA:2024hlh}, as they strongly indicate the need to remove \emph{all} resolvable GBs before searching for EMRIs. 

Can the EMRI signal corrupt the detection of GBs? The answer is ``no''. An individual EMRI signal has very weak overlap and small matched-filtering SNR with any individual GB. The observed high false-detection $\rho$ is due to the \emph{net positive cumulative} effect of interactions between the EMRI and all resolvable GBs. We confirmed this by performing Bayesian inference of a few bright GBs in the presence of an EMRI signal. Essentially, GBs do not ``see'' the EMRI signal. However, the situation might change if the LISA data contained $\gtrsim 1000$ EMRIs simultaneously.

We should also mention an important caveat. This work is a proof of principle, and we do not present it as a comprehensive analysis of the problem. We limited our analysis to \emph{one} particular EMRI and a rather narrow prior range of parameters. We specifically chose an EMRI designed to maximize its overlap with the Galactic population in the frequency domain. The effect of resolvable GBs should be reduced if the central MBH is either heavier ($\sim 10^7\,M_\odot$) or lighter ($\sim 10^5\,M_\odot$). However, the EMRI considered here represents one of the most anticipated type of system based on recent theoretical studies (Fig. 3.6 of~\cite{LISA:2024hlh}, see also~\cite{Qunbar:2023vys, Mancieri:2024sfy}). 

\begin{acknowledgments}
All simulations were performed on the CC-IN2P3 cluster. We thank Ollie Burke for very useful discussions. S.B. acknowledges support from the CNES for the exploration of LISA science and funding from the French National Research Agency (grant ANR-21-CE31-0026, project MBH waves). 

This work made use of packages \texttt{corner}~\cite{corner}, \texttt{matplotlib}~\cite{Hunter:2007},  \texttt{numpy}~\cite{harris2020array} and \texttt{scipy}~\cite{2020SciPy-NMeth}.

\end{acknowledgments}

\clearpage

\bibliographystyle{apsrev4-2}
\bibliography{bib_emri}

\clearpage

\appendix
\section{Full posterior}
\label{appendix}

Here we give a posterior distribution for all 11 variable parameters of the EMRI signal in the ``Full-Galaxy'' run. As before, the main results given by orange contours are compared to the results of the ``Noiseless'' run shown in blue.

\vspace{\textfloatsep} 

\noindent\makebox[\textwidth][c]{%
  \begin{minipage}{\dimexpr2\columnwidth+\columnsep\relax}%
    \centering
    \includegraphics[width=\linewidth]{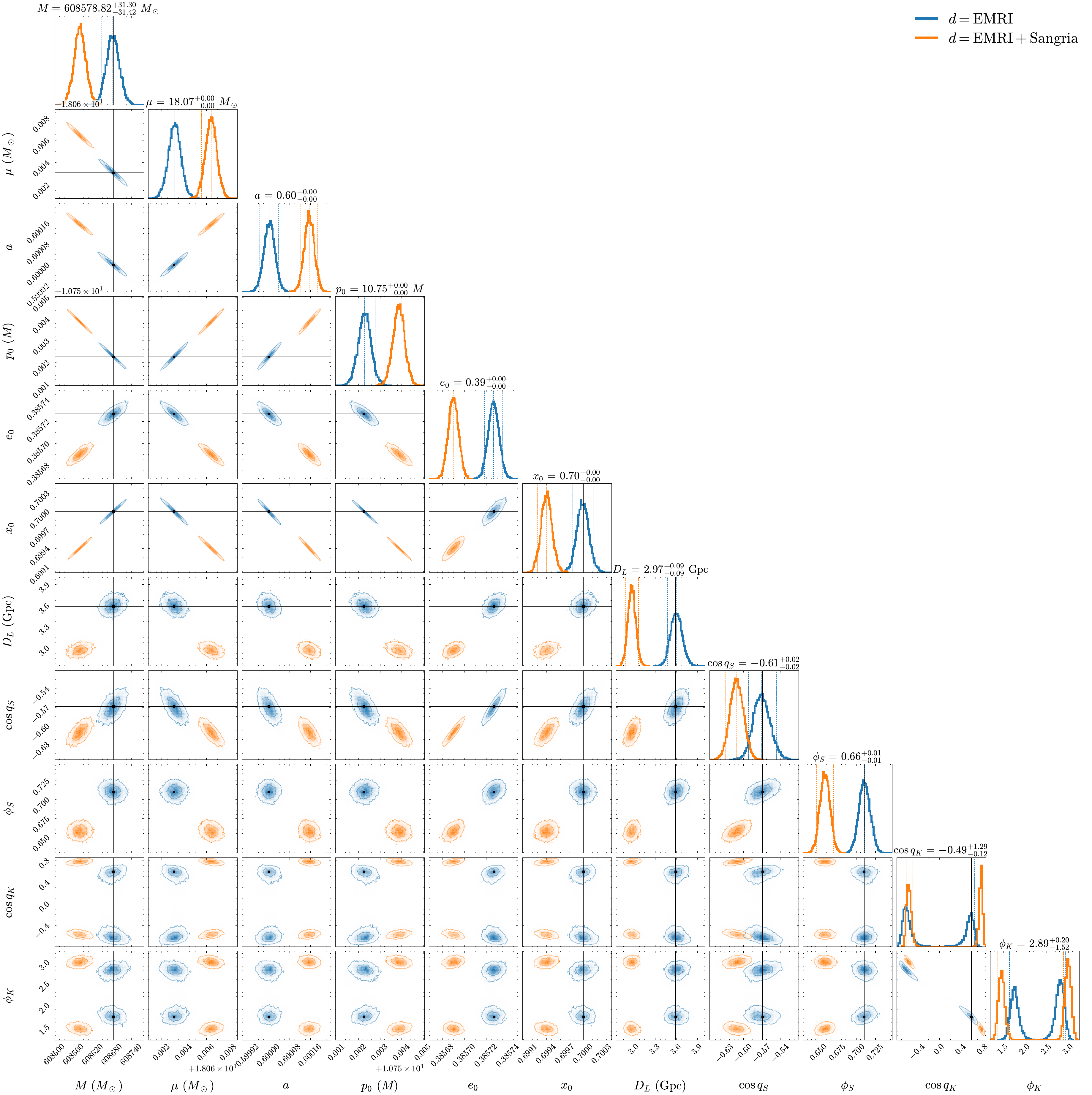}\\[6pt]
    \captionof{figure}{\centering Influence of the full population of Galactic binaries on EMRI parameter estimation. Corner plot shows Bayesian inference of the 11 variable waveform parameters. The true point is marked in black. Contour lines mark the 50\% and 90\% credible regions. The posteriors for the noiseless run are given in blue for comparison.}
    \label{fig:full_main_mcmc}
  \end{minipage}%
}%

\clearpage

\end{document}